\documentclass[manuscript, letterpaper]{aastex6}
\usepackage{graphicx}
\usepackage[suffix=]{epstopdf}
\usepackage{natbib}
\usepackage{amsmath}
\usepackage{url}
\usepackage{xspace}

\makeatletter 
\long\def\frontmatter@title@above{
\vspace*{-\headsep}\vspace*{\headheight}
\noindent\footnotesize
{\noindent\footnotesize\textsc{\@journalinfo}}\par
{\noindent\scriptsize Preprint typeset using \LaTeX\ style AASTeX6\\
With modifications by David W. Hogg, Daniel Foreman-Mackey, Boris Leistedt
}\par\vspace*{-\baselineskip}\vspace*{0.625in}
}%
\makeatother
\makeatletter
\let\origsection\section
\renewcommand\section{\@ifstar{\starsection}{\nostarsection}}
\newcommand\nostarsection[1]{\sectionprelude\origsection{#1}}
\newcommand\starsection[1]{\sectionprelude\origsection*{#1}}
\newcommand\sectionprelude{\vspace{1em}}
\let\origsubsection\subsection
\renewcommand\subsection{\@ifstar{\starsubsection}{\nostarsubsection}}
\newcommand\nostarsubsection[1]{\subsectionprelude\origsubsection{#1}}
\newcommand\starsubsection[1]{\subsectionprelude\origsubsection*{#1}}
\newcommand\subsectionprelude{\vspace{1em}}
\makeatother
\sloppypar

\definecolor{mycolor}{RGB}{0,104,139}
\hypersetup{colorlinks,breaklinks,
            urlcolor=mycolor,
            citecolor=mycolor,
            linkcolor=mycolor}
            
\journalinfo{Prepared for ApJ}

\newcommand{\ie}{{\textit{i.e.},~}}
\newcommand{\eg}{{\textit{e.g.},~}}
\newcommand{\equref}[1]{{\xspace}Eq.~(\ref{#1})}
\newcommand{\figref}[1]{{\xspace}Fig.~\ref{#1}}
\newcommand{\figrefs}[2]{{\xspace}Figs.~\ref{#1}~and ~\ref{#2}}

\newcommand{\secref}[1]{{\xspace}Sec.~\ref{#1}}
\renewcommand{\d}{{\mathrm{d}}}
\newcommand{\equ}[1]{\begin{equation}#1\end{equation}}
\newcommand{\eqn}[1]{\begin{eqnarray}#1\end{eqnarray}}

\newcommand{\nobj}{{N_{\rm stars}}}

\begin{document}

\title{Hierarchical probabilistic inference\\ of the color-magnitude diagram\\ and shrinkage of stellar distance uncertainties}
  
\shorttitle{Probabilistic inference of the color--magnitude diagram}
\shortauthors{Leistedt et al}

\author{
	Boris~Leistedt\altaffilmark{1,2},
	David~W.~Hogg\altaffilmark{1,3,4}
	}

  \altaffiltext{1}{Center for Cosmology and Particle Physics, Department of Physics, \\ New York University, 726 Broadway, New York, NY 10003, USA}
  \altaffiltext{2}{NASA Einstein Fellow}
  \altaffiltext{3}{Center for Data Science, New York University, 60 Fifth Avenue, New York, NY 10011, USA}
  \altaffiltext{4}{Flatiron Institute, 162 Fifth Avenue, New York, NY 10010, USA}

\begin{abstract}
We present a hierarchical probabilistic model for improving geometric stellar distance estimates using color--magnitude information. 
This is achieved with a data driven model of the color--magnitude diagram, not relying on stellar models but instead on the relative abundances of stars in color--magnitude cells, which are inferred from very noisy magnitudes and parallaxes.
While the resulting noise-deconvolved color--magnitude diagram can be useful for a range of applications, we focus on deriving improved stellar distance estimates relying on both parallax and photometric information.
We demonstrate the efficiency of this approach on the 1.4 million stars of the Gaia TGAS sample that also have APASS magnitudes.
Our hierarchical model has 4~million parameters in total, most of which are marginalized out numerically or analytically.
We find that distance estimates are significantly improved for the noisiest parallaxes and densest regions of the color--magnitude diagram. 
In particular, the average distance signal-to-noise ratio and uncertainty improve by 19~percent and 36~percent, respectively, with 8~percent of the objects improving in SNR by a factor greater than 2.
This computationally efficient approach fully accounts for both parallax and photometric noise, and is a first step towards a full hierarchical probabilistic model of the Gaia data.
\end{abstract}

\keywords{stars: distances, stars: C--M diagrams, methods: statistical.} 

\newpage
\section{Introduction}

Gaia and LSST will soon provide parallaxes for many millions of stars inhabiting our Galaxy \citep{gaia, lsstsciencebook}.
However, parallax measurements for most of these objects are very noisy due to parallax errors significantly increasing with distance.
Most analyses adopt signal-to-noise cuts, thus discarding large amounts of hard-won data. 
In rare cases, uncertainties are explicitly accounted for, via probabilistic hierarchical models, for example \citep[\eg][]{sesar2016}.
Regardless of how parallax or distance uncertainties are accounted or ignored, reducing them can improve a variety of studies and increase the scientific return of a mission like Gaia.

In addition to direct geometric determination via parallaxes, stellar distances can be indirectly inferred from spectral information.
This is because the vast majority of stars live in compact regions of color--magnitude space, such as the main sequence or the red giant branch.
This information can be used as a prior for disentangling the contributions of distance and absolute magnitude in apparent magnitudes.
In other words, photometric information can complement parallax information and prove crucial to determine distances.
The influence of distance priors, as well as color--magnitude priors based on stellar models, was explored in previous work \citep[\eg][]{Astraatmadja22016, Astraatmadja2016}, which relied on theoretical models or simulations.
Yet, stellar models predict narrow tracks in color--magnitude, which are reductive of the diversity of real objects, and may introduce biases when used as priors for determining distances.
In this paper, we address this problem and present a purely data-driven approach for improving parallax-based distance estimates with color--magnitude information inferred directly from the data, with no use of external models. 

Our approach relies on a standard property of hierarchical probabilistic models: by using a large number of noisy observations (here, parallaxes and magnitudes) one can construct an estimate of the underlying distribution (here, a color--magnitude diagram), which is in turn applied as a prior to shrink the uncertainties on the  properties of the individual objects (here, the distances).
This \textit{Bayesian shrinkage} naturally occurs within hierarchical models when computing or sampling the full joint posterior distribution. 
In other words, \textit{the whole is greater than the sum of the parts}, and stellar distance estimates can be improved by simultaneously analyzing all the available data.
This is particularly relevant for Gaia, which will soon provide many millions of sources with parallax, proper motion, and photometric measurements, offering the opportunity to construct accurate models of the distribution of magnitudes, colors, dust, and stellar densities in the Galaxy, with minimal external data. 
Those models will in turn significantly improve our estimates of the intrinsic properties of individual stars.
The core goal of this paper is to demonstrate how one aspect of this hierarchical inference can be done in practice. 
We focus on evaluating the shrinkage of distance estimates obtained by constructing a flexible color--magnitude diagram based on the joint analysis of millions of noisy parallaxes and apparent magnitudes.
We make a number of simplifying assumptions to make our demonstration more transparent, such as adopting uninformative distance priors and point estimates of dust corrections. 
Those assumptions could be relaxed or complexified in subsequent work. 

In parallel to this work, a noise-deconvolved color--magnitude diagram reconstruction from the Gaia data with a variant of the extreme deconvolution (XD) algorithm \cite{bovy2011} is presented in \cite{Anderson2017}.
Five key elements distinguishes the XD inference from the present work.
First, the XD model is based on 2MASS photometry, which is more precise than the APASS data we use here.
Second, in the XD model, the color--magnitude diagram is modeled as a (transformed) Gaussian mixture where the positions and sizes of the mixture components are also inferred.
Third, in the XD model, the noise model is more approximate (magnitude uncertainties are ignored) to make the method more tractable.
Fourth, in the XD model dust is a smaller correction (because of the infrared photometry) and it is iteratively optimized to be self-consistent with the final color--magnitude diagram model.
Fifth, the XD model is a maximum-marginalized likelihood method; it is not fully probabilistic at all levels of the hierarchy in the hierarchical model.
Those differences allow to construct a color--magnitude diagram at higher resolution, at the cost of an approximate methodology scheme \citep{Anderson2017}.
The methodology and results presented here fully account for a realistic noise model (for both parallaxes and photometry), but they rely on noisier data and keep the resolution (the positions and widths of the Gaussian mixture component) fixed, resulting in a more reliable inference, but of a less precise color--magnitude diagram.
Suitably combining the two approaches will be the key to fully exploiting the full-Mission Gaia data with computationally tractable but nonetheless justified hierarchical probabilistic models.

This paper is structured as follows: our model and inference framework are described in \secref{sec:model}, and applied to the Gaia data in \secref{sec:application}. 
Conclusions and perspectives are presented in \secref{sec:concl}.

\section{Model}\label{sec:model}

\begin{table} 
\centering
\begin{tabular}{cl}
\hline
$s$	&	object index (the $s$-th star)\\
$d_s, \varpi_s, M_s, C_s$	&	true distance, parallax, absolute magnitude, and color	\\
$\hat{\varpi}_s, \sigma_{\hat{\varpi}_s}^2$ 	&	parallax estimate and its variance\\
$\hat{m}_s, \hat{C}_s, \sigma^2_{\hat{m}_s}, \sigma^2_{\hat{C}_s}$ 	&	apparent magnitude and color estimates, and their variances\\
$b_s$	&	index of the color--magnitude bin of the $s$th object\\
\hline
$b$	&	generic index of color--magnitude bin (the $b$th bin)\\
$n_b$	& 	object count in the $b$-th color--magnitude bin  \\
$\{n_b\}$	&	set of all galaxy counts $n_b$, summing to $\nobj$\\
$f_b$	&	fractional galaxy count in the $b$-th color--magnitude bin  \\
$\{f_b\}$	&	set of all fractional bin counts $f_b$, summing to $1$\\
$\{ d_s, b_s\}$	&	distances and bins of all stars in the sample	\\
$\{ \hat{m}_s, \hat{C}_s \}$ &	all magnitude and color estimates\\
\hline
\end{tabular}
\caption{Summary of our notation. }
\label{tab:notation}
\end{table} 

We consider a set of stars indexed as $s=1, \cdots, \nobj$, each characterized by a (latent) distance $d_s$, absolute magnitude $M_s$, and color $C_s$. 
We consider only one magnitude and one color for simplicity, but it should be noted that the model and method presented below can be straightforwardly extended to multiple magnitudes and colors.

Intrinsic properties like distance and absolute magnitude are not observable, they must be inferred from apparent magnitudes and parallax measurements.  
The estimate of the parallax is denoted $\hat{\varpi}_s$ and is assumed to have a Gaussian variance $\sigma_{\hat{\varpi}_s}^2$.
We will consider two magnitudes, $\hat{m}_s$ and $\hat{m}^\prime_s$, assumed to be uncorrelated and to have Gaussian variances $\sigma_{\hat{m}_s}^2$ and $\sigma_{\hat{m}^\prime_s}^2$.
We will use the first one $\hat{m}_s$ as a reference magnitude for inferring the absolute magnitude $M_s$, and the second one to form a color estimate $\hat{C}_s =\hat{m}^\prime_s - \hat{m}_s $ with Gaussian variance $\sigma_{\hat{C}_s}^2 = \sigma_{\hat{m}_s}^2 + \sigma_{\hat{m}^\prime_s}^2$.

We aim at estimating the distance $d_s$ of each star from the noisy data $\hat{m}_s$,  $\hat{C}_s$ and $\hat{\varpi}_s$. 
While distance is directly connected to the parallax via $\varpi_s=1/d_s$, it is also informed by the apparent magnitude since $m_s = M_s + 5\log_{10} d_s$ where $d_s$ is expressed in units of $10$ pc.
Note that when only the apparent magnitude is available, distance and absolute magnitude are degenerate and cannot be disentangled. 
This degeneracy is partially broken with the parallax information.
Here, we seek to incorporate the knowledge that stars do not have arbitrary colors and magnitude.
The way this information enters distance estimates is made obvious by writing the posterior probability distribution on the distance,
\eqn{
	p(d_s | \hat{m}_s, \hat{C}_s, \hat{\varpi}_s) = \int \d M_s \ \d C_s \ p\bigl(\hat{m}_s, \hat{C}_s, \hat{\varpi}_s \bigr\rvert M_s, d_s, C_s\bigr) \ p\bigl( M_s, d_s, C_s \bigr) \label{eq:naivedistposterior}.
}
This integral marginalizes over the true absolute magnitude and color.
This might be expensive to perform numerically, but the choices we will make below will allow us to execute it analytically.

The first term of \equref{eq:naivedistposterior} is a likelihood function, and the second term is the prior. 
Assuming that the magnitude and parallax estimates are independent, the likelihood function factorizes as the product of two terms, 
\equ{
	p\left(\hat{\varpi}_s \bigr\rvert d_s\right) = \mathcal{N}\bigl(\hat{\varpi}_s - d_s^{-1};\sigma_{\hat{\varpi}_s}^2 \bigr),\label{eq:parallaxlike}
}
and
\equ{
	p\bigl(\hat{m}_s, \hat{C}_s \bigr\rvert M_s, d_s, C_s\bigr)  =  \mathcal{N}\bigl( M_s + 5\log_{10}d_s  -\hat{m}_s ;\sigma_{\hat{m}_s}^2 \bigr) \  \mathcal{N}\bigl(\hat{C}_s - C_s;\sigma_{\hat{C}_s}^2 \bigr).
}

The final term, $ p( M_s, d_s, C_s ) $, is the prior knowledge about the distances, magnitudes, and colors of stars. 
It is typically factorized as $p( M_s, C_s | d_s) p (d_s)$, where $p (d_s)$ is the distance prior, which could depend on position and based on a three-dimensional model of the Galaxy. 
The color--magnitude prior is typically made distance-independent, $p( M_s, C_s )$, and based on stellar models \citep[\eg][]{Astraatmadja22016, Astraatmadja2016}. 

In this work, we will adopt a uniform distance prior and focus on the magnitude--color term, which we parametrize as $p(M_s, C_s  \bigr\rvert \{ f_{b} \} ) $.
We construct a model of the relative abundance of objects in color--magnitude cells (\ie in two dimensions).
The color--magnitude distribution is described as a linear mixture of $B$ components,
\equ{
	p\left(M_s, C_s  \bigr\rvert \{ f_{b} \} \right) = \sum_{b=1}^B f_b \ K_b(M_s, C_s),\label{eq:cmd}
} 
with $K_b$ the  $b$th kernel. 
The parameters $\{ f_{b} \}$ refer to the relative probabilities of finding objects in the various kernels, and must sum to one ($\sum_b f_b = 1$).

For the kernels, we adopt Gaussian distributions to make the integral of \equref{eq:naivedistposterior} analytically tractable.
The $b$-th kernel will be centered at $(\mu_{b,0}, \mu_{b,1})$ and have a diagonal covariance $(\sigma_{b,0}^2, \sigma_{b,1}^2)$.
We take $\mu_{b+1,0}-\mu_{b,0} = \sigma_{b,0}$ and $\sigma_{b,0}$ constant (similarly for the color dimension) to uniformly and contiguously tile a rectangular region of interest in color--magnitude space. 
With this parameterization, the integral of \equref{eq:naivedistposterior} is tractable and leads to
\eqn{
	p(d_s &&| \hat{m}_s, \hat{C}_s, \hat{\varpi}_s, \{ f_{b} \})  \ \propto  \ \sum_b\ f_b \ \mathcal{N}\bigl(\hat{\varpi}_s - d_s^{-1};\sigma_{\hat{\varpi}_s}^2 \bigr) \\ 
	 \quad\quad \times \ \ &&\mathcal{N}\bigl( \mu_{b_s,0} + 5\log_{10}d_s  -\hat{m}_s ;\sigma_{\hat{m}_s}^2 + \sigma_{b_s,0}^2 \bigr)   \ \mathcal{N}\bigl(\hat{C}_s - \mu_{b_s,1};\sigma_{\hat{C}_s}^2 + \sigma_{b_s,1}^2 \bigr). \label{eq:distposterior}\nonumber
}

Finally, to facilitate parameter inference, we will introduce a latent variable $b_s$ denoting the bin the $s$th object belongs to.
Then, we can equivalently write the color--magnitude model as
\eqn{
	p\left(b_s \bigr\rvert \bigl\{ f_b \bigr\}\right) \ &=& \ f_{b_s} \\ 
	p\left(M_s, C_s \bigr\rvert b_s \right) \ &=& \ \mathcal{N}\bigl(M_s - \mu_{b,0};\sigma_{b,0}^2 \bigr)  \ \mathcal{N}\bigl(C_s - \mu_{b,1};\sigma_{b,1}^2 \bigr).\nonumber
}
Our notation is summarized in Table~\ref{tab:notation}.
Note that in all of the above we have assumed that all magnitudes are properly dereddenned, \ie that the absorption by interstellar dust has been corrected for.
As discussed below, we will use the distance point estimate $1/\hat{\varpi}_s$ for performing this correction.
While this is formally incorrect (since the reddenning correction depends on distance which is a parameter we are inferring), this approximation will be sufficient for the case-study of this paper.

\subsection{Inference}

Since we fix the kernel locations $\{  (\mu_{b,0}, \mu_{b,1}) \}$ and covariances $\{(\sigma_{b,0}^2, \sigma_{b,1}^2)\}$, our color--magnitude model is fully described by the relative amplitudes $\{ f_{b} \}$. 
Those could be set by prior knowledge (\eg external data or stellar models), in which case one could use \equref{eq:distposterior} to infer the distance of each object.
Here, we seek to infer $\{ f_{b} \}$ from the data.
Thus, the full posterior of interest is $p(\{ d_s \}, \{ f_{b} \} | \{ \hat{m}_s, \hat{C}_s, \hat{\varpi}_s \})$, which has $B + \nobj$ parameters.

Given the number of parameters and the natural degeneracies between magnitudes and distances, standard sampling techniques may be difficult to apply.
However, the inference is greatly simplified by focusing on $p(\{ b_s \}, \{ f_{b} \} | \{ \hat{m}_s, \hat{C}_s, \hat{\varpi}_s \})$, where distances are marginalized over.
We adopt a Gibbs sampling strategy \citep[see \eg][]{Casella1992, Wandelt2004, Levin2009, brooks2011handbook}, which consists of alternating draws from the conditional distributions of $\{ b_s \}$ and $\{ f_{b} \}$.
At the $i$th iteration, we will draw new values of the $\{ f_{b} \}$ and $\{b_s\}$ parameters given the values of the previous iteration.
In other words, we first draw $\{ f_{b} \}^{(i)}$ given $\{b_s\}^{(i-1)}$ (and the data), and for each object then draw $b_s^{(i)}$ given $\{ f_{b} \}^{(i)}$ (and the data). 
The sequence $\{ f_{b} \}^{(i)},\{b_s\}^{(i)}$ for $i=1, \cdots, N_\mathrm{samples}$ forms a Markov Chain with the target posterior distribution of interest as equilibrium distribution.
This allows us to avoid the magnitude--distance degeneracies and exploit other properties of our model, such as conjugate priors.
We now detail how to draw from the correct conditional distributions.

The first draw is fairly standard: with the bin locations $\{b_s\}$ fixed, the fractional amplitudes $\{ f_{b} \}$ follow a Dirichlet distribution entirely determined by $\{n_b \}$, with $n_b$ the number of objects in the $b$-th bin.
All the other parameters enter the constant proportionality factor, so the target conditional distribution is
\eqn{
	p\left(\bigl\{ f_b \bigr\} \bigr\rvert \bigl\{ d_s, b_s, \hat{m}_s, \hat{C}_s, \hat{\varpi}_s \bigr\} \right) \ = \ p\bigl( \bigl\{ f_b \bigr\} \bigr\rvert \{n_b \} \bigr) \ \propto\  \prod_b \frac{ f_b^{n_b} }{n_b !},
}
where we have assumed the simplest uninformative prior for the relative amplitudes, \ie that they are positive and normalized to one, $p(\bigl\{ f_b \bigr\}) = \delta^D(1 - \sum_b f_b) \prod_b \Theta(f_b)$, where $\delta^D$ is the Dirac delta function and $\Theta$ the step function.
Given the number counts $\{n_b \}$, one can draw $\{ f_b \}$ from this distribution using standard tools for Dirichlet distributions. 
Alternatively, one could draw $\gamma_b \sim \mathrm{Gamma}(n_b + 1)$ and take $f_b = \gamma_b / \sum_b' \gamma_b'$.

The second step of the Gibbs sampler is to draw the bins $\{b_s\}$ given the amplitudes $\{f_b\}$. 
Those draws can be performed independently (thus, in parallel) over objects.
This may not look simple since we have to marginalize over the distances, and the target conditional distribution to sample is
\eqn{
	p\left(b_s \bigr\rvert \bigl\{ f_b \bigr\}, \hat{m}_s, \hat{C}_s, \hat{\varpi}_s\right) \ &\propto& \  \int p\left(b_s, d_s \bigr\rvert \bigl\{ f_b \bigr\}, \hat{m}_s, \hat{C}_s, \hat{\varpi}_s\right)  \mathrm{d} d_s.
}  
However, closer examination reveals that the bins can be simply drawn from a multinomial distribution with amplitudes given by
\eqn{
	p\left(b_s \bigr\rvert \bigl\{ f_b \bigr\}, \hat{m}_s, \hat{C}_s, \hat{\varpi}_s\right) \ &\propto& \ f_{b_s}  g_{b_s}
} 
with
\eqn{
	g_{b_s} &=& \int \mathcal{N}\bigl( \mu_{b_s,0} + 5\log_{10}d_s  -\hat{m}_s ;\sigma_{\hat{m}_s}^2 + \sigma_{b_s,0}^2 \bigr)    \\ && \times \ \mathcal{N}\bigl(\hat{C}_s - \mu_{b_s,1};\sigma_{\hat{C}_s}^2 + \sigma_{b_s,1}^2 \bigr) \ \mathcal{N}\bigl(\hat{\varpi}_s - d_s^{-1};\sigma_{\hat{\varpi}_s}^2 \bigr)\  \mathrm{d}  d_s.
}
Since we keep the kernel locations and covariances fixed, those weights can be precalculated via $\nobj$ independent dimensional integrals.
We evaluate them numerically given that the integrand is simple and the integration can be tuned easily. 
If this operation were prohibitive, or if the kernel parameters were not kept fixed, one could sample the distance $d_s$ (jointly or conditionally on the bin $b_s$).
Since the kernels and likelihoods admit simple gradients and Hessians, one could use Hamiltonian Monte Carlo \citep[\eg][]{Duane1987, Neal2012} to efficiently sample all $\nobj$ distances simultaneously.

\subsection{Discussion}

\begin{figure}
\hspace*{-3mm}\includegraphics[width=15.75cm]{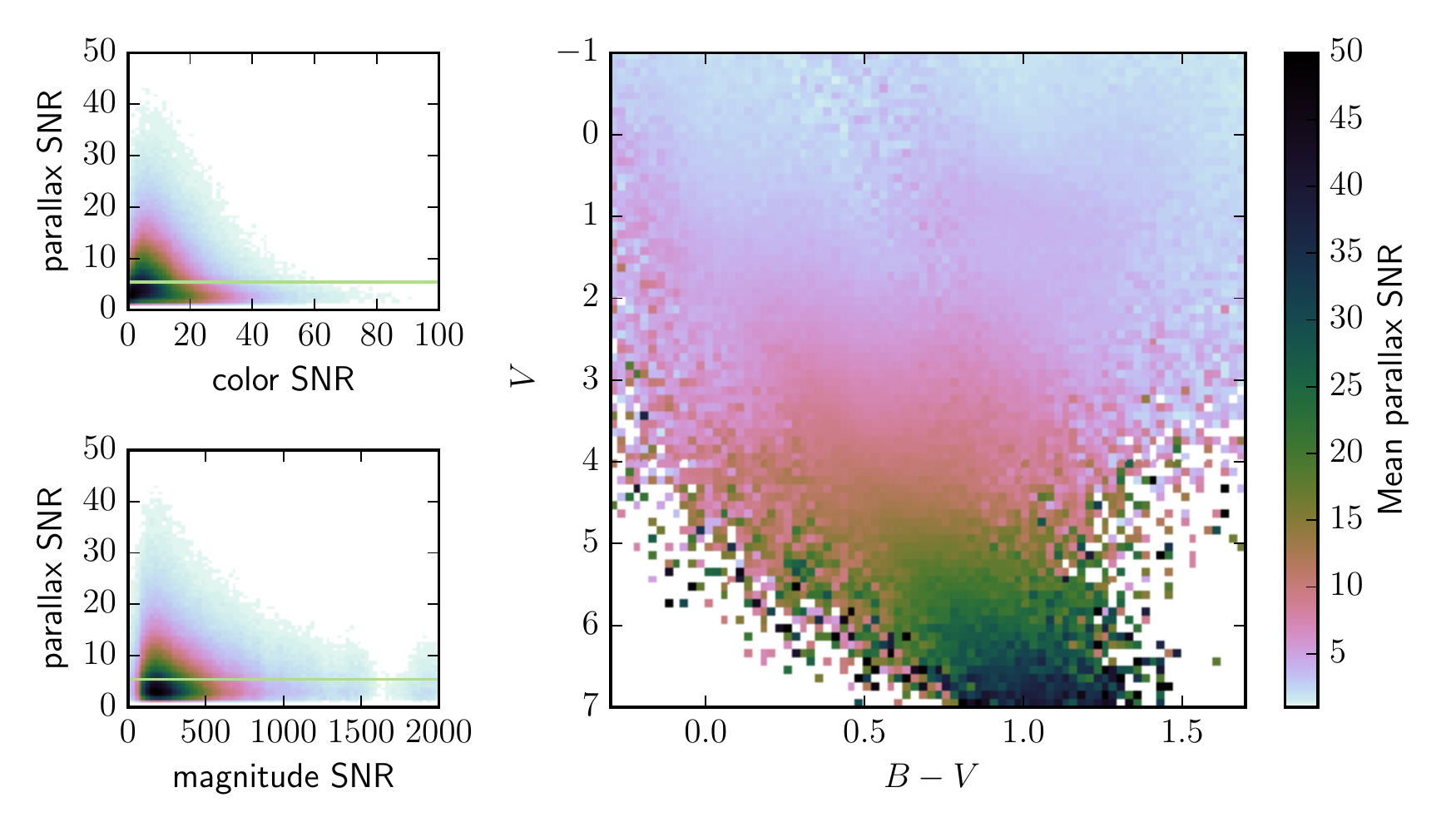}
\caption{Distributions of the magnitude, color, and parallax signal-to-noise ratios (SNR) of the Gaia TGAS+APASS data we fit and validate our model on. The line indicates the parallax SNR level used to split the data into two sub-samples containing the `best' and `worst' parallaxes. The right panel shows the average parallax SNR in color--magnitude cells, illustrating how the upper part of the color--magnitude diagram is dominated by low-SNR objects. Reconstructing the noiseless color--diagram requires an inference framework capable of correctly dealing with uncertainties in colors, magnitudes, and parallaxes.}
\label{fig:datasummary}
\end{figure}

We now briefly discuss some of the advantages and limitations of our approach.

First, we restricted our attention to the color--magnitude diagram, and neglected any dependency on other quantities such as galactic latitude.
We also adopted uniform distance priors.
This can be corrected by adopting or constraining models of the 3D distribution stellar densities, described by kernel mixtures, for example \citep[see \eg][]{Astraatmadja22016, Astraatmadja2016}.  
Although this extension is technically trivial, we have not developed it since we focus on how color--magnitude information informs distance estimates without the use of stellar models.
Similarly, our framework could be extended to other observables such as velocities.

Second, our kernel mixture model offers a significant amount of freedom for describing the color--magnitude diagram, but could be improved in various ways.
Changing the kernels would not strongly affect our inference framework, unless they were not differentiable, in which case it would not possible to use Hamiltonian Monte Carlo.
Analytic marginalization of true magnitudes and colors was made possible by adopting Gaussian kernels. 
Furthermore, we did not optimized the positions and covariances of the kernels, unlike in standard Gaussian Mixture models \citep[\eg][]{bovy2011, Anderson2017}. 
Compared to those, our tiling of color--magnitude space requires more components (many of which are zero) but is easy to initialize, and also converges quickly. 

Third, we assumed that the magnitudes are perfectly dust-corrected. 
However, dust extinction depends on distance, which is a parameter of our model.
Furthermore, reliable 3D dust maps are only available for a limited region of three-dimensional space.
In principle, dust corrections should really be inferred jointly with the absolute magnitudes and colors of the data at hand. 
This is explored in \cite{Anderson2017}.

\section{Application to Gaia TGAS}\label{sec:application}

\begin{figure}
\hspace*{-3mm}\includegraphics[width=15cm]{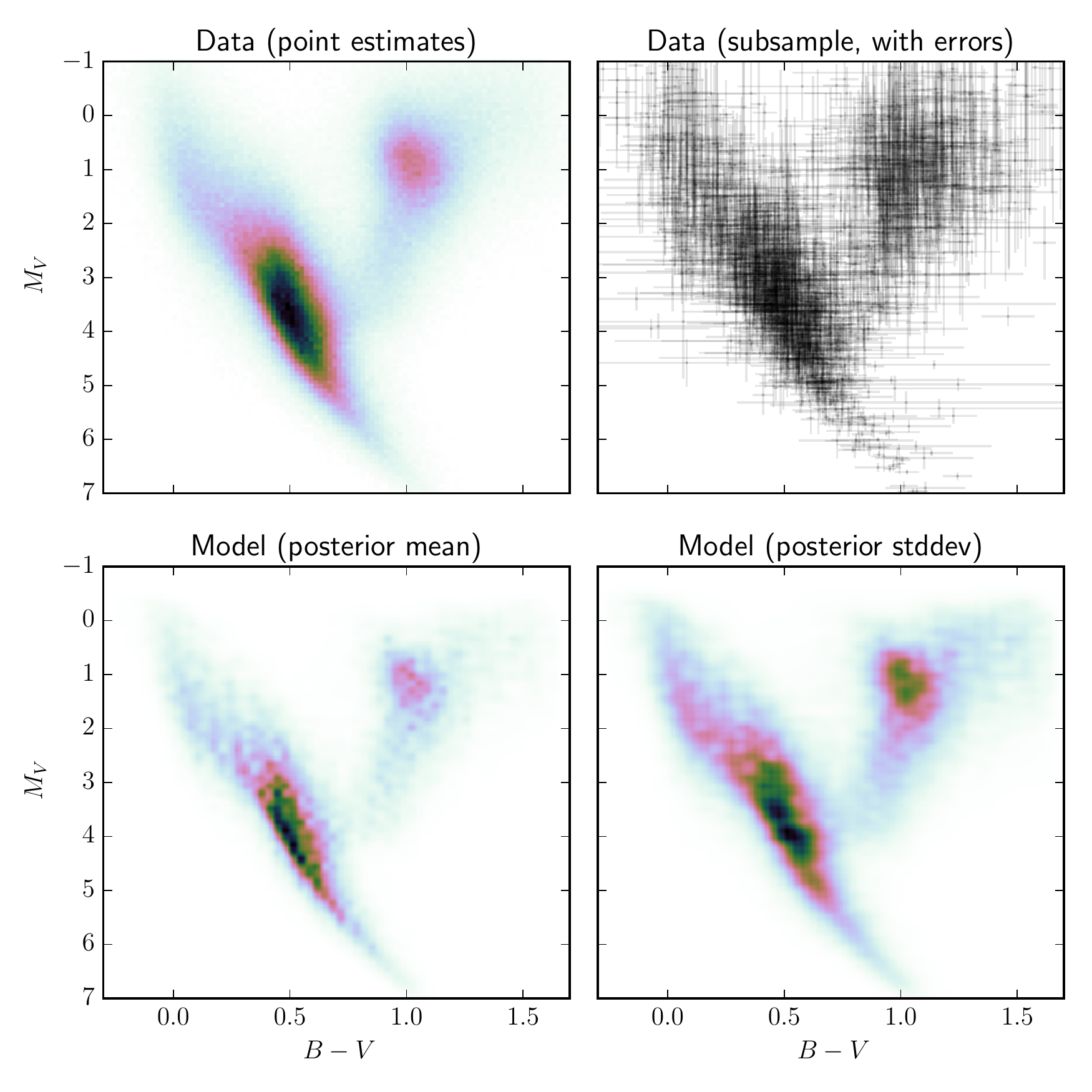}
\caption{Upper left panel: color--magnitude diagram based on the noisy data, obtained with magnitude and parallax point estimates (left). Upper right panel: two thousand objects randomly selected from the main data sample. We show the approximate (linearized) Gaussian errors resulting from the parallax and photometric noise (our framework correctly deals with those errors with no approximation). Lower left and right panels: mean and standard deviation of our model, which is the result of deconvolving all observational errors of the data shown in the upper panels and in \figref{fig:datasummary} into a noiseless color--magnitude diagram described as a mixture of Gaussians tiling the color--magnitude region of interest. }
\label{fig:colmagdiag_mainsample}
\end{figure}

\begin{figure}
\hspace*{-3mm}\includegraphics[width=15.5cm]{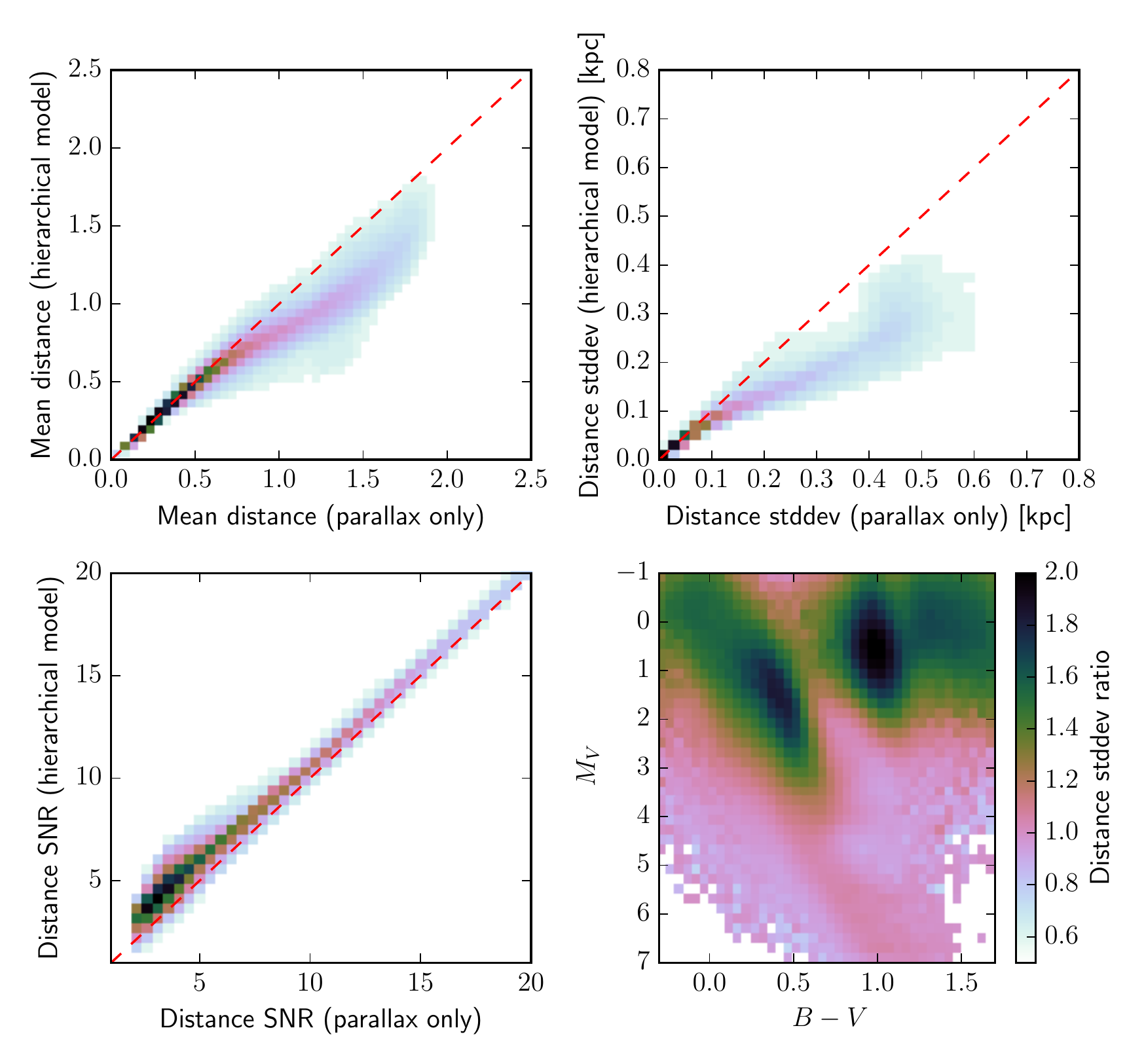}
\caption{Distances obtained when sampling the hierarchical model which produced the color--diagram shown in \figref{fig:colmagdiag_mainsample}. 
The first three panels shows the change in the mean, standard deviation, and SNR of the distance estimate (based on the posterior distribution), with the number counts in logarithmic scale. The final panel shows the ratio of standard deviations placed in the color--diagram (standard method over hierarchical model). The shrinkage of the uncertainties is a consequence of the hierarchical nature of the model, and is most efficient for low-SNR objects and the densest parts of the color diagram.}
\label{fig:colmagdiag_mainsample_dist}
\end{figure}

\begin{figure}
\hspace*{-2mm}\includegraphics[width=15.7cm]{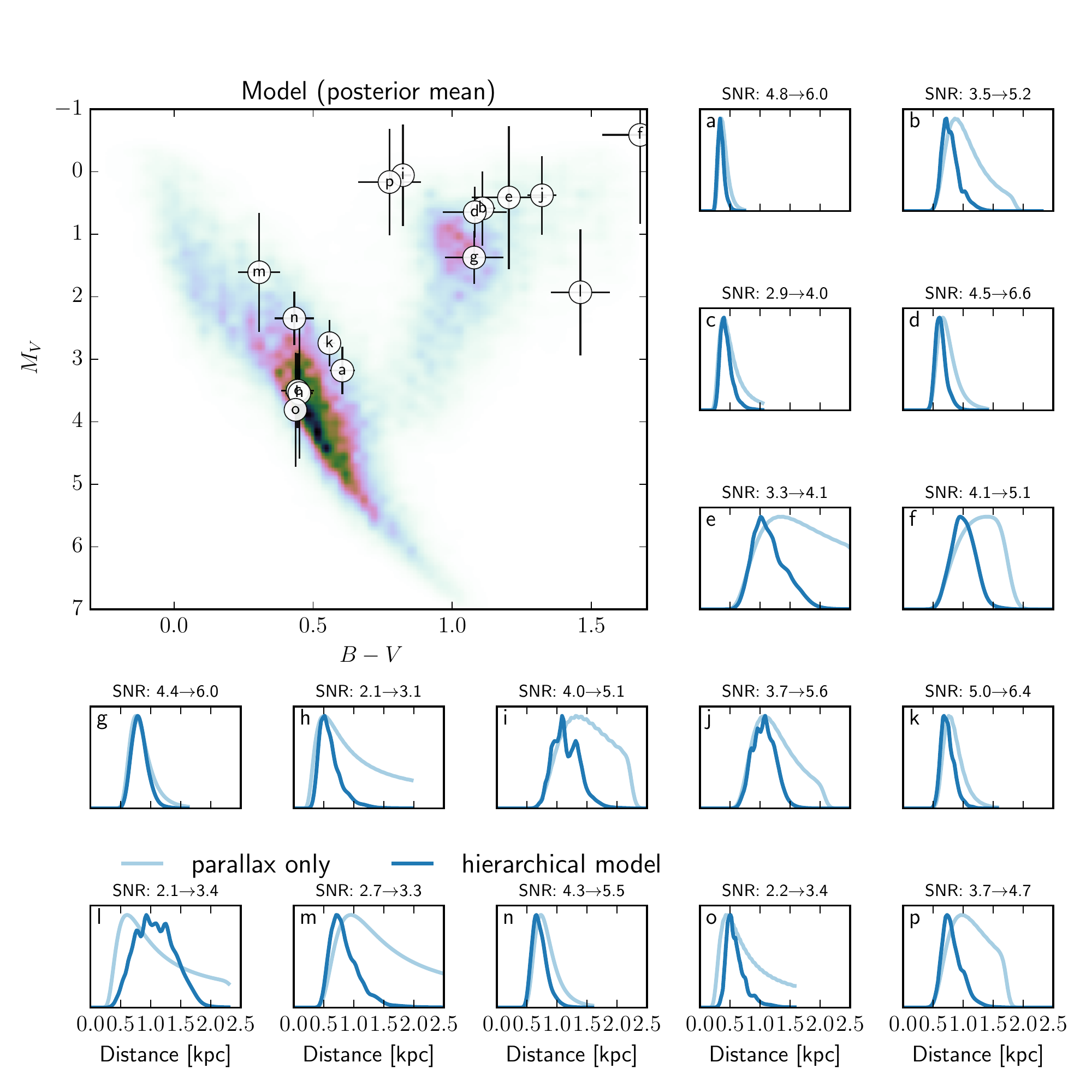}
\caption{Posterior distributions on the distances of a few objects involved in constraining the model shown in \figrefs{fig:colmagdiag_mainsample}{fig:colmagdiag_mainsample_dist}. 
The improvement in distance SNR is also shown. The objects are also placed on the inferred color--magnitude diagram, to highlight that the shrinkage is most efficient for low-SNR objects and the densest parts of the color diagram.}\label{fig:model_dist_pdfs}
\end{figure}

We consider the Gaia data \citep{gaia}, specifically the first data release (DR1) of the Tycho-Gaia astrometric solution \citep[hereafter TGAS,][]{gaia_dr1}.
We restrict our attention to the objects with valid B and V magnitudes from the AAVSO Photometric All Sky Survey (APASS) Data Release 9 \citep{munari2014, hendenmunari2014}. 
We also remove objects with parallax signal-to-noise ratio (SNR) lower than 1. 
This leads to 1.4 million objects with magnitude, color, and parallax information. 
We do not apply more stringent parallax or color cuts since the purpose of our method is exactly to construct a color--magnitude model from both low- and high-SNR objects. 
Finally, we apply dust corrections based on position and distance point estimate ($1/\hat{\varpi}$) with the three-dimensional dust map of \cite{Green2015bayestar}. 
Our data sample is summarized in \figref{fig:datasummary}, which shows the magnitude, color, and parallax SNR distributions.
The bulk of the objects has parallax SNR lower than 10 and is at $M_V  < 4$, in the upper part of the color--magnitude diagram.
This highlights the need for a correct inference framework that exploits all objects, since focusing on high-SNR objects would bias the results and prevent us from correctly inferring the fainter regions of the color--magnitude space.
This is illustrated below.
Note that we did not add the recommended parallax error offset of 0.3 \citep{gaia_dr1} because overestimated errors typically cause excessive deconvolution.
For the type of hierarchical inference we are considering, underestimated errors lead to more conservative results.

We create a validation sample by extracting 10\% of the objects, adding significant amount of noise to the parallax estimates, and verifying that our framework improves the distances consistently with the original values. 
Further details about this process are given below.
We also split the main sample according to parallax SNR, into two samples of equal size containing the `best' and `worst' parallaxes. 
We perform the inference on those two samples as well as their union.
 
For each of the three samples, we use the Gibbs sampler presented above to draw thirty thousand samples of the fractional amplitudes and bins. 
Thus, our model has $B + \nobj$ sampled parameters, and $3 \nobj$ parameters marginalized over (the true magnitudes, colors, and distances), either numerically or analytically.
The positions and widths of the kernels are obtained by uniformly tiling (with 60 points) the color and magnitude lines in the ranges $[-0.3, 1.7]$ and $[-1, 7]$, respectively.

The top panels of \figref{fig:colmagdiag_mainsample} show the input data. 
In particular, the right panel shows a random sample of two thousand objects with their associated errors.
Note that the absolute magnitude Gaussian error shown for each object is approximate and obtained by linearizing the parallax likelihood and adding the resulting error to the photometric one in quadrature.
This is formally incorrect (and only adopted for visualization purposes) since it underestimates the true absolute magnitude error.
Our framework does not make this approximation since it explicitly models those components.

The mean and standard deviation of the resulting color--magnitude diagram (with bins marginalized over) are shown in the bottom panels of \figref{fig:colmagdiag_mainsample}.
Those were obtained by computing the mean and standard deviation of the set of models obtained by applying \equref{eq:cmd} to all the posterior samples we drew.   
As expected, the models recovered are significantly narrower than the data since we are effectively deconvolving observational errors to produce a noiseless color--magnitude diagram. 
The main sequence, its turn-over, and the giant branch, are successfully recovered at higher resolution than in the input noisy data.
Features corresponding to binary stars and red clump stars are not clearly visible due to the coarse binning we adopted. 
While we could increase the resolution of our model, there is no guarantee that those features will be resolved since the photometric noise is significant in our data.
Adopting Gaia and 2MASS photometry could help resolve this issue \citep[see][]{Anderson2017}.
In addition, our Gibbs sampling scheme is sensitive to the resolution adopted, and the mixing of parameters slows down as the resolution increases.
One could circumvent this obstacle by adopting a Hamiltonian Monte Carlo sampling scheme.
This is possible since the gradients with respect to the parameters are simple and can be efficiently calculated.
Finally, we note that the oscillatory features are also the result of an excessive resolution of the model in the noisiest regions of color--magnitude space.
This could be avoided by adopting variable kernel size (inferred or tuned a priori).

\figref{fig:colmagdiag_mainsample_dist} shows a comparison of the stellar distances with and without the extra information brought by the color--magnitude model.
For each object we compute the distance posterior distribution via \equref{eq:naivedistposterior} on a grid, and we average over samples of the bin allocations and relative amplitudes of the color--magnitude model, which is equivalent to marginalizing them out.
Our point of comparison is the set of parallax-only distance estimates, \ie not using our hierarchical model but geometric information only, via \equref{eq:parallaxlike}.
Note that neither the full nor the marginalized posterior distributions are Gaussian, as expected and also shown below. 
Nevertheless, the mean and standard deviation provides a useful metrics. 
Our inference methodology provides samples of the full, joint non-Gaussian posterior distribution, and those samples should be used for any subsequent inference relying on the color--magnitude diagram to correctly capture the data and model uncertainties.
We measure that on average the distance SNR improves by 19\% and the distance uncertainty decreases by 36\%.
We find that 8\% of the objects have their distance uncertainty halved after the inclusion of color--magnitude information via our hierarchical model. 
This shrinkage of the distance uncertainties is most efficient in the densely populated regions of color--magnitude space.
\figref{fig:model_dist_pdfs} shows the distance posterior distributions obtained with our method for a few randomly chosen objects, illustrating the uncertainty shrinkage.
We also place these objects in the color--magnitude diagram; the errors on the absolute magnitude are obtained by resampling.
Note that as in \figref{fig:colmagdiag_mainsample}, this underestimates the absolute magnitude uncertainty, but nevertheless provides a qualitatively useful error estimate.

\begin{figure}
\hspace*{-4mm}\includegraphics[width=15.9cm]{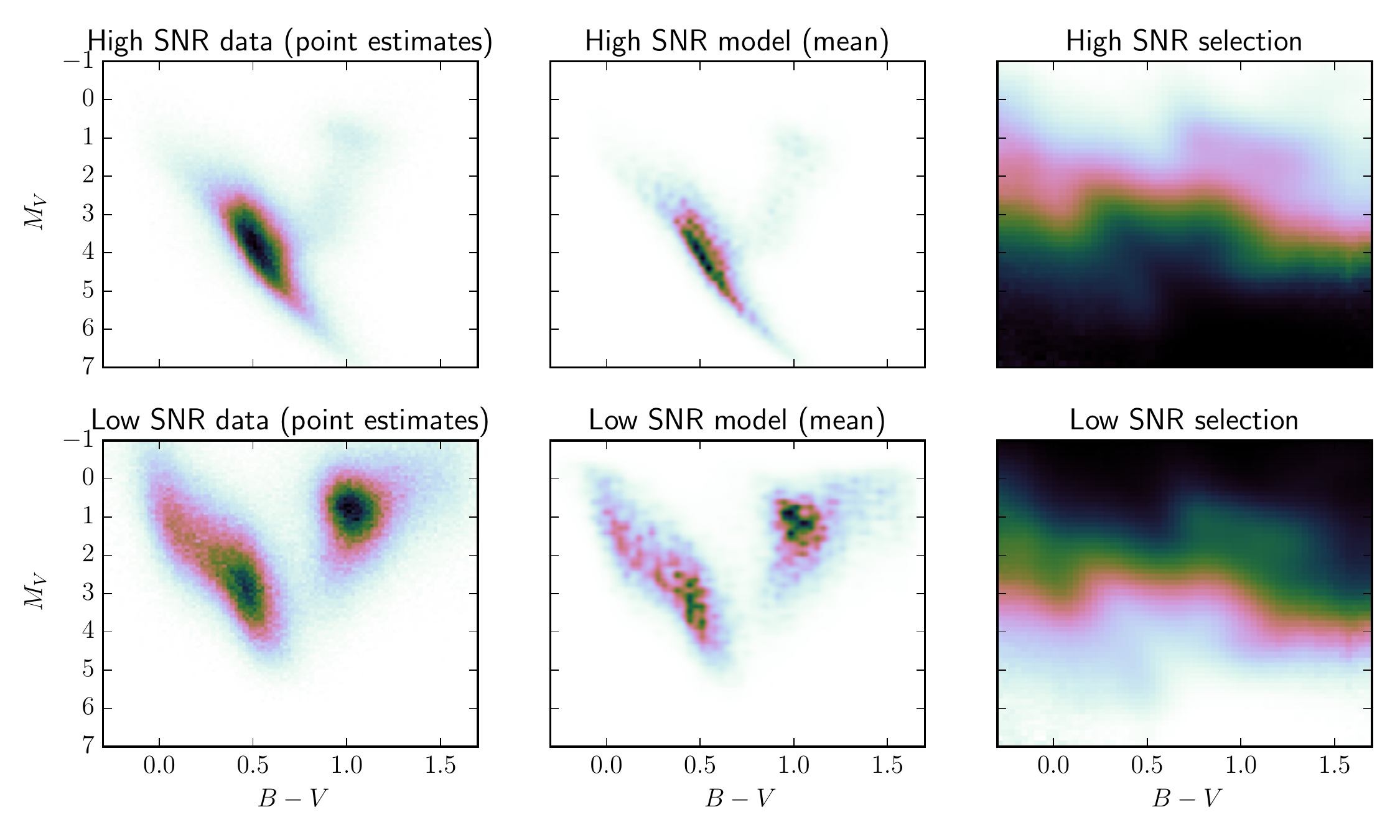}
\caption{Same as \figref{fig:colmagdiag_mainsample} for with the main sample split based on parallax SNR. This highlights the contributions of the stars with the `best' and `worst' parallaxes to the color--magnitude diagram, and the importance of using a correct scheme for inferring the latter in the presence of significant observational errors. The right panels show the probability of detecting sources in both subsamples (approximated by resampling the data), which demonstrates how one misses a significant fraction of color--magnitude space by applying parallax SNR cuts. It is by using all of the objects, not only high-SNR ones, that a satisfactory color--magnitude diagram can be inferred.}
\label{fig:colmagdiag_othersamples}
\end{figure}

\figref{fig:colmagdiag_othersamples} shows the mean and standard deviation of the color--magnitude diagram resulting from performing the inference on the subsamples with parallax SNR cuts (\ie splitting our main sample at parallax SNR of 5). 
Those demonstrate that including the noisiest objects is essential for correctly inferring the faint regions of magnitude space.
The main sequence is well recovered with the high-SNR objects, while the red giant branch is barely detected. 
By contrast, it is well recovered with the low-SNR objects, but the main sequence is then partially erased.
This is a natural consequence of the SNR increasing with absolute magnitude.
This highlights the importance of a correct probabilistic framework, capable of correctly exploiting data with heterogeneous noise to reconstruct the noiseless color--magnitude diagram.
To illustrate the effect of the SNR cut we also resampled the data and kept objects that satisfied the cuts.
This provides a first-order estimate of the detection probability or selection function of those two subsamples, shown in the right panels of \figref{fig:colmagdiag_othersamples}, illustrating what regions of the color--magnitude diagram are unexplored.

\begin{figure}
\hspace*{-3mm}\includegraphics[width=15.8cm]{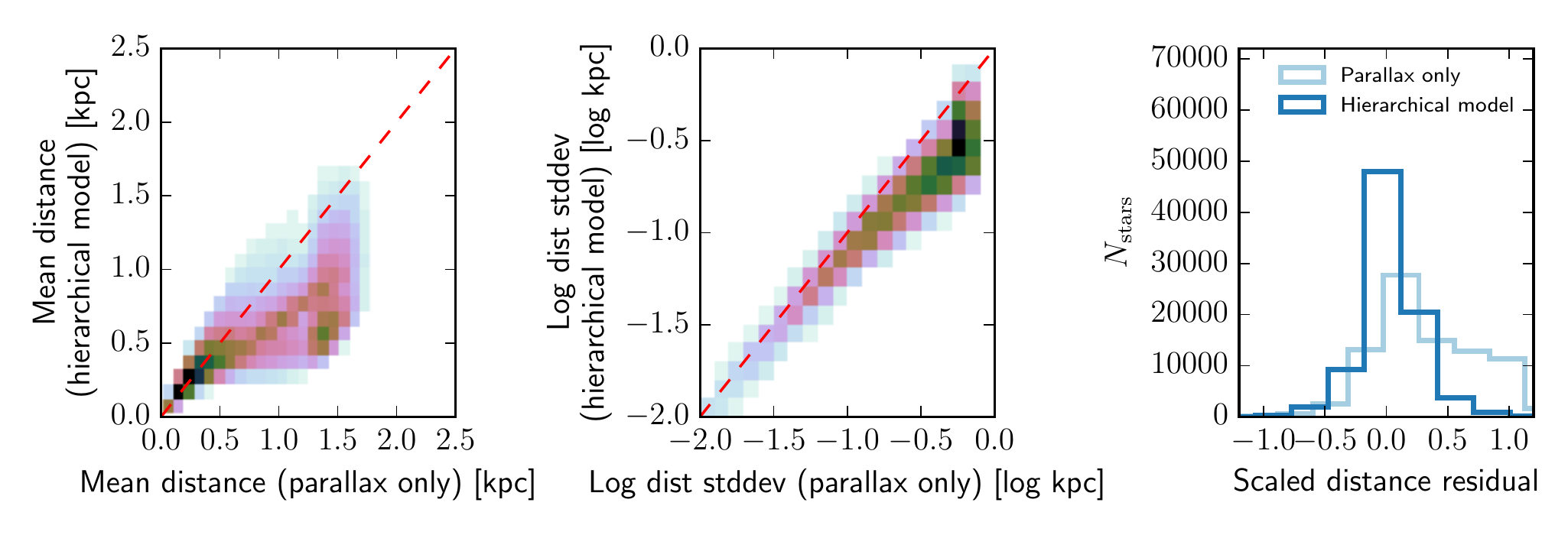}
\caption{Mean, standard deviation, and scaled residuals (truth minus mean estimate, divided by standard deviation) of the distances in our validation sample, based on the posterior distributions.
Given the more significant levels of noise the distance are more significantly improved than in our main sample. 
The mean residuals are not zero due to the non-Gaussianity of the posterior distributions.}
\label{fig:cv_metrics}
\end{figure}

We now turn to the validation sample. 
Since we do not know the true distances for most objects in our data set, we add noise to the parallax estimate, at a level equal to ten times the parallax error. 
We then compute the posterior distribution on the distance (on a distance grid) with and without color--magnitude information, as in \figref{fig:model_dist_pdfs}. 
We simply use the mean model shown in \figref{fig:colmagdiag_mainsample} as a color-magnitude prior. 
The results are shown in \figref{fig:cv_metrics}.
Given that those objects have significant amounts of noise, causing the distance posterior distribution to be highly non-Gaussian, the mean distance overestimates the true distance (the original parallax-based estimate). 
The hierarchical model significantly decreases this effect, \ie improves the distance estimates both in terms of mean and uncertainty, demonstrating the validity of our inference scheme.

Finally, we perform an additional test of our method on open clusters.
We retrieve the coordinates and distances of known open clusters from the WEBDA database\footnote{www.univie.ac.at/webda}.
From our TGAS-APASS sample, we select the objects within a radius corresponding to 3 pc around each cluster, using the known cluster distance. 
We only keep objects with parallaxes consistent with the open cluster distance at the three sigma level, to reduce the number of candidate members.
\figref{fig:cv_metrics} shows the result of applying our framework to those objects for a few selected clusters, \ie using the color--diagram inferred above to inform the distance estimates.
Even though not all objects selected by this procedure are true cluster members, most distance estimates undergo a visible improvement toward the cluster distance, with the SNRs mildly improved.
In addition, \figref{fig:model_dist_pdfs} showed that the distance posterior distributions produced by the hierarchical model are significantly more Gaussian, \ie the typical heavy tail produced by the parallax-only likelihood function for low SNR objects is strongly decreased.
This approach could potentially help identifying cluster members and determining cluster distances with minimal external information.

\begin{figure}
\hspace*{-3mm}\includegraphics[width=15.5cm, trim = 0cm 1.6cm 0cm 0.6cm, clip]{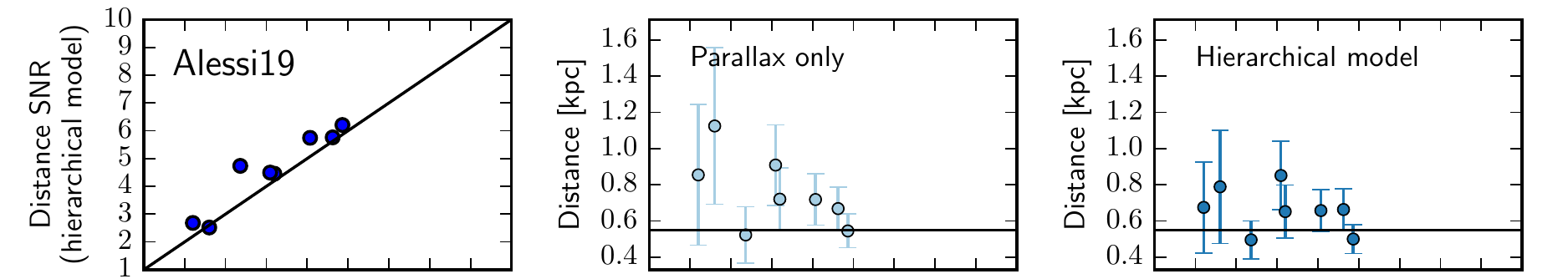}
\hspace*{-3mm}\includegraphics[width=15.5cm, trim = 0cm 1.6cm 0cm 0.6cm, clip]{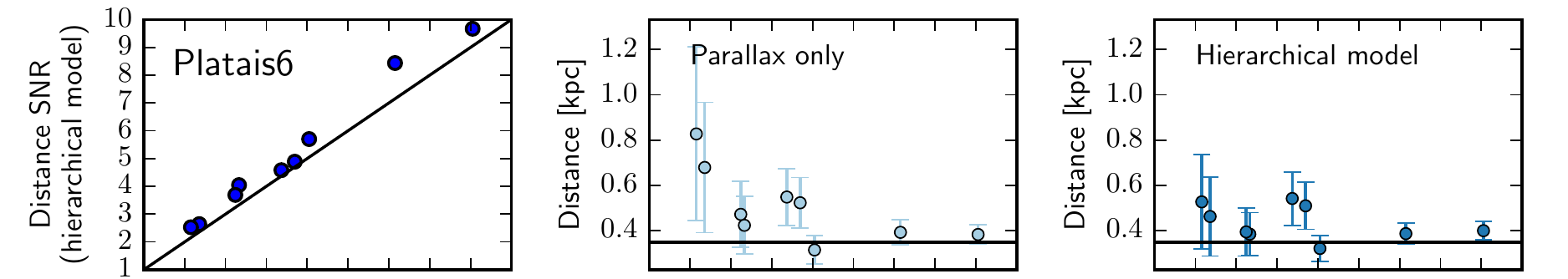}
\hspace*{-3mm}\includegraphics[width=15.5cm, trim = 0cm 1.6cm 0cm 0.6cm, clip]{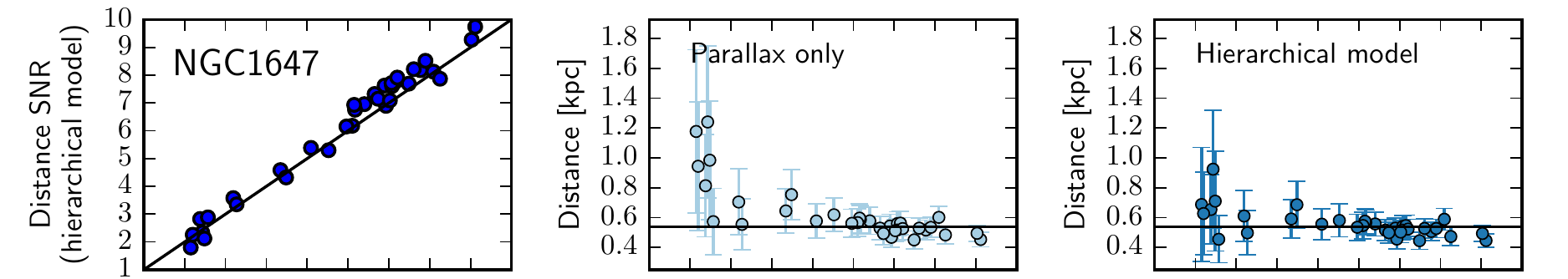}
\hspace*{-3mm}\includegraphics[width=15.5cm, trim = 0cm 1.6cm 0cm 0.6cm, clip]{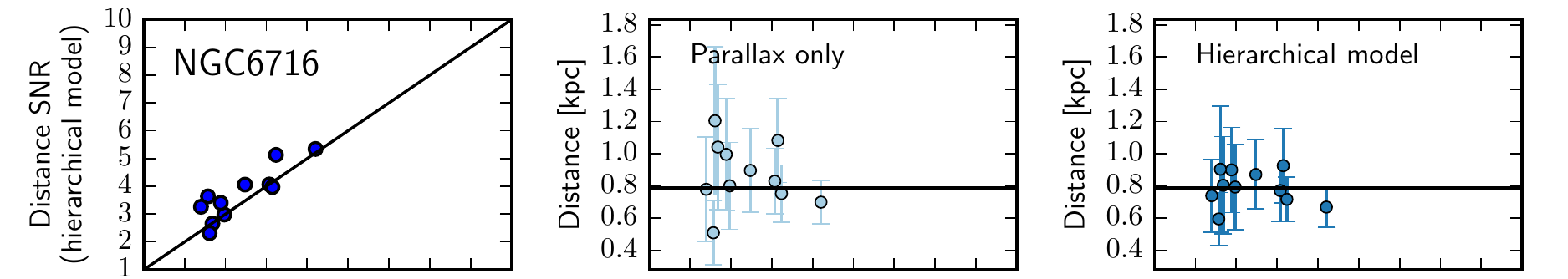}
\hspace*{-3mm}\includegraphics[width=15.5cm, trim = 0cm 1.6cm 0cm 0.6cm, clip]{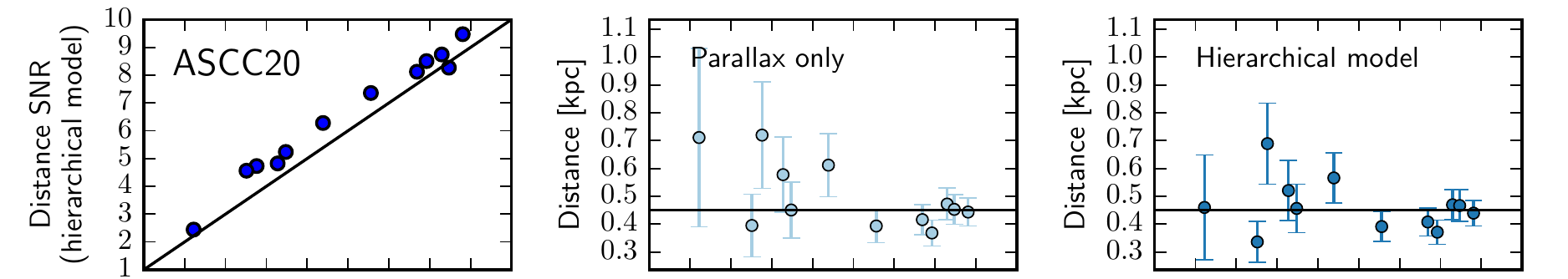}
\hspace*{-3mm}\includegraphics[width=15.5cm, trim = 0cm 0cm 0cm 0.6cm, clip]{NGC6716_metrics}
\caption{Distances estimates of candidate members of a few open clusters in our data set, demonstrating the shrinkage of the distance uncertainties, consistently with the known distance to the open cluster.}
\label{fig:oc_metrics}
\end{figure}

\newpage\vspace*{1cm}
\section{Conclusions}\label{sec:concl}

Stellar distances are ubiquitous in astronomy and are most directly determined geometrically via parallaxes. 
However, parallax measurements are unreliable for the faintest, most distant objects, which inhabit most of the Galaxy. 
We presented a framework for obtaining improved distance estimates with both parallax and color--magnitude information without external data or priors such as stellar models. 
This exploits a well-known property of probabilistic hierarchical models: by utilizing the wealth of information concealed in millions of data points (the noisy and less noisy ones), one can model the underlying distributions and in turn improve the quality of the individual parameter estimates. 
We presented a version of this idea where the color--magnitude distribution of stars is directly inferred from noisy parallax and magnitude measurements via an efficient parameter inference scheme.
This is the first color--magnitude model consistently capturing both parallax and photometric uncertainties.
We applied this methodology to Gaia+APASS data and demonstrated that it leads to significant improvements in the distance estimates, specifically for the faintest, most distant objects.

The framework described here included a number of simplifying assumptions. 
First, we used uniform distance priors and described the color--magnitude diagram as a mixture of Gaussian kernels with fixed positions and widths.
Those two assumptions could easily be relaxed without affecting our inference methodology or the quantitative results presented here.
Second, we ignored selection effects that may distort the color--magnitude diagram and the distance distribution.
Those could be incorporated but require external data and modifications to the sampling method.
Third, we neglected the dependency of the color--magnitude on other properties (\eg galactic latitude), and we performed the magnitude reddenning corrections using existing three-dimensional dust maps evaluated at the parallax-based distance point estimates.
Our framework could also support those extensions with moderate technical changes.
However, in order to fully take advantage of the available data, one should jointly \textit{infer} the three-dimensional distribution of dust and stellar density, and compare with results based on physical models and priors.
Any selection effects modulating the detection probability or the measured properties or uncertainties of stars should also be included.
This requires more substantial technical developments due to the large number of interlaced parameters.
We intend to explore this avenue in future work.

\acknowledgments

We thank Lauren Anderson, Andy Casey, Jo Bovy, Adrian Price-Whelan, Semyeong Ho, Keith Hawkins, David Spergel, and the Stars Group Meeting at the Flatiron Institute, for stimulating discussions and useful comments on this manuscript.

BL was supported by NASA through the Einstein Postdoctoral Fellowship (award number PF6-170154).
DWH was partially supported by the NSF (AST-1517237) and the Moore--Sloan Data Science Environment at NYU.

This project was developed in part at the 2016 NYC Gaia Sprint, hosted by the Center for Computational Astrophysics at the Simons Foundation in New York City.

This work has made use of data from the European Space Agency (ESA) mission Gaia (\url{http://www.cosmos.esa.int/gaia}), processed by the Gaia Data Processing and Analysis Consortium (DPAC, \url{http://www.cosmos.esa.int/web/gaia/dpac/consortium}). Funding for the DPAC has been provided by national institutions, in particular the institutions participating in the Gaia Multilateral Agreement.

This research was made possible through the use of the AAVSO Photometric All-Sky Survey (APASS), funded by the Robert Martin Ayers Sciences Fund.


\bibliography{bib}

\end{document}